# Artificial Neural Network Modeling of the Conformable Fractional Isothermal Gas Spheres

Yosry A. Azzam[1], Emad A.-B. Abdel-Salam[2] and Mohamed I. Nouh[1]

[1]Astronomy Department, National Research Institute of Astronomy and Geophysics (NRIAG), 11421 Helwan, Cairo, Egypt

[2]Department of Mathematics, Faculty of Science, New Valley University, El-Kharja 72511, Egypt

e-mail: mohamed.nouh@nriag.sci.eg

**Abstract:**

The isothermal gas sphere is a particular type of Lane–Emden equation and is used widely to model many problems in astrophysics like stars, star clusters, and galaxies' formations. In this paper, we present a computational scheme to simulate the conformable fractional isothermal gas sphere using an artificial neural network (ANN) technique and compare the obtained results with the analytical solution deduced using the Taylor series. We performed our calculations, trained the ANN, and tested it using a wide range of the fractional parameter. Besides the Emden functions, we calculated the mass-radius relations and the density profiles of the fractional isothermal gas spheres. The results obtained provided that ANN could perfectly simulate the conformable fractional isothermal gas spheres.

1. Introduction

In the last decade, fractional differential equations played a very important role in science and engineering. One of the most interesting fractional differential equations arising in physics, astrophysics, engineering, and chemistry, is the Lane-Emden equations and Emden-Fowler equations. Many methods are proposed to solve these equations. The fractional polytropic models were investigated by El-Nabulsi (2011) for the white dwarf stars, Bayin and Krisch (2015) for the incompressible gas sphere, Abdel-Salam and Nouh (2016) and Yousif et al. (2021) for the isothermal gas sphere. Analytical solutions to the fractional Lane-Emden equations using series expansion and Adomian decomposition methods are introduced by Nouh and Abdel-Salam (2018a), Abdel-Salam and Nouh (2020), Nouh and Abdel-Salam (2018b), Abdel-Salam et al. (2020).

Artificial Neural Networks (ANN) are proved to be a very promising tool that has been used in wide areas of scientific issues and have found many applications to solve problems related to geophysics, engineering, environmental sciences, and astronomy fields, Weaver (2000), Tagliaferri et al. (1999), Tagliaferri and Longo (2003), Faris et al. (2014), Elminir et al. (2007), El-Mallawany et al. (2014), Leshno et al. (1993), Lippmann (1989), Zhang (2000). Besides, ANNs have found several works for modeling many physical phenomena and have been used widely to solve fractional and integer differential equations problems that used different patterns for ANN architecture Raja et al. (2010), Raja et al. (2011), Raja et al. (2015), Hadian-Rasanan et al. (2020), Pakdaman et al. (2017), Zúñiga-Aguilar et al. (2017). In addition, Ahmad et al. (2017) used the artificial neural networks (ANNs) to compute the solution of Lane–Emden-type equations. Recently, Nouh et al. (2020) presented a solution to the fractional polytropic gas sphere (first kind of the Lane-Emden equation), the results indicated that using the ANN method is precise when compared with other methods.

Considering the isothermal equation of state given by

$$p = K\rho, \tag{1}$$

where $K$ is the pressure constant. By implementing the principles of the conformable derivatives Khalil et al. (2014, Yousif et al. (2021) derived the conformable second-order nonlinear differential equation that describes the isothermal gas sphere as

$$\frac{1}{x^{2\alpha}} \frac{d^\alpha}{d x^\alpha} \left( x^{2\alpha} \frac{d^\alpha u}{d x^\alpha} \right) = e^{-u} \tag{2}$$

The mass contained in the sphere is given by

$$M(x^\alpha) = 4\pi \left[ \frac{K}{4\pi G} \right]^{\frac{3}{2}} \rho_c^{-\frac{3}{2}} \left[ \left( x^{2\alpha} \frac{d^\alpha u}{d x^\alpha} \right) \right], \tag{3}$$

the radius is given by

$$R^\alpha = \left[ \frac{K}{4\pi G} \right]^{\frac{1}{2}} \rho_c^{-\frac{1}{2}} x^\alpha \tag{4}$$

and the density is given by

$$\rho = \rho_c \, e^{-u}, \tag{5}$$

In the current work, we shall solve the fractional isothermal gas sphere equation using the Taylor series and train the ANN algorithm by using tables of the fractional Emden functions, mass-radius relations, and density profiles. For the sake of computational simulation, the normal feed-forward neural network was used to approximate the fractional Emden function solution, mass-radius relations, and density distributions which gives good agreement when compared to other analytical schemes. The architecture used in this research is a feed-forward neural network that has three-layers and is trained using the back-propagation algorithm based on the gradient descent rule. The rest of the paper is organized as follows: Section 2 deals with the Taylor expansion solution of the fractional isothermal gas sphere equation. The mathematical modeling of the neural network is performed in section 3. In section 4, the results are introduced with discussions. The conclusion reached is given in section 5.

## 2. Taylor expansion of Equation (2)

We used the principles of the conformable fractional derivatives developed by Khalil et al. (2014) to solve the fractional isothermal gas sphere. Write Equation (2) as

$$D_x^{\alpha\alpha}u + \frac{2\alpha}{x^\alpha}D_x^\alpha u - e^{-u} = 0, \quad u(0) = 0, \; D_x^\alpha u(0) = 0, \tag{6}$$

The fractional Taylor series solution for any function $u(x)$ can be written as

$$u(x) = u(0) + \frac{D_x^\alpha u(0)}{\alpha}x^\alpha + \frac{D_x^{\alpha\alpha}u(0)}{2!\alpha^2}x^{2\alpha} + \frac{D_x^{\alpha\alpha\alpha}u(0)}{3!\alpha^3}x^{3\alpha} + \frac{D_x^{\alpha\alpha\alpha\alpha}u(0)}{4!\alpha^4}x^{4\alpha} + \frac{D_x^{\alpha\alpha\alpha\alpha\alpha}u(0)}{5!\alpha^5}x^{5\alpha} + ....,$$

Equation (6) can be written in the following form

$$x^\alpha D_x^{\alpha\alpha}u = -2\alpha D_x^\alpha u + x^\alpha e^{-u} \tag{7}$$

Differentiate Equation (6) with respect to $\alpha$ we get

$$\begin{aligned}\alpha D_x^{\alpha\alpha}u + x^\alpha D_x^{\alpha\alpha\alpha}u &= -2\alpha D_x^{\alpha\alpha}u + \alpha e^{-u} - x^\alpha e^{-u}D_x^\alpha u, \\ \Rightarrow 3\alpha D_x^{\alpha\alpha}u + x^\alpha D_x^{\alpha\alpha\alpha}u &= \alpha e^{-u} - x^\alpha e^{-u}D_x^\alpha u,\end{aligned} \tag{8}$$

Put $x = 0$ in the last equation we have

$$3\alpha D_x^{\alpha\alpha}u(0) = \alpha e^0, \qquad \Rightarrow D_x^{\alpha\alpha}u(0) = \frac{1}{3},$$

Differentiating Equation (8) with respect to $\alpha$ we have

$$4\alpha D_x^{\alpha\alpha\alpha}u + x^\alpha D_x^{\alpha\alpha\alpha\alpha}u = -2\alpha e^{-u}D_x^\alpha u + x^\alpha e^{-u}(D_x^\alpha u)^2 - x^\alpha e^{-u}D_x^{\alpha\alpha}u, \tag{9}$$

When $x = 0$ we have

$4\alpha D_x^{\alpha\alpha\alpha} u = 0, \qquad \Rightarrow D_x^{\alpha\alpha\alpha} u = 0,$

Differentiating Equation (9) with respect to $\alpha$ we have

$5\alpha D_x^{\alpha\alpha\alpha\alpha} u + x^\alpha D_x^{\alpha\alpha\alpha\alpha\alpha} u = 3\alpha e^{-u}(D_x^\alpha u)^2 - x^\alpha e^{-u}(D_x^\alpha y)^3 + 3x^\alpha e^{-u} D_x^\alpha u D_x^{\alpha\alpha} u - 3\alpha e^{-u} D_x^{\alpha\alpha} u - x^\alpha e^{-u} D_x^{\alpha\alpha\alpha} u,$

When $x = 0$ we have

$5\alpha D_x^{\alpha\alpha\alpha\alpha} u = -3\alpha e^{-u} D_x^{\alpha\alpha} u, \Rightarrow D_x^{\alpha\alpha\alpha\alpha} u(0) = -\dfrac{3}{5} e^0 \left(\dfrac{1}{3}\right) = -\dfrac{1}{5},$

and so on. Finally, we have

$u(x) = u(0) + \dfrac{D_x^\alpha u(0)}{\alpha} x^\alpha + \dfrac{D_x^{\alpha\alpha} u(0)}{2!\alpha^2} x^{2\alpha} + \dfrac{D_x^{\alpha\alpha\alpha} u(0)}{3!\alpha^3} x^{3\alpha} + \dfrac{D_x^{\alpha\alpha\alpha\alpha} u(0)}{4!\alpha^4} x^{4\alpha} + ....$ ,

So the solution of Equation (6) is given by

$$u(x) = \dfrac{1}{6\alpha^2} x^{2\alpha} - \dfrac{1}{120\alpha^4} x^{4\alpha} + .... \qquad (10)$$

### 3. Neural network algorithm

### 3.1. Mathematical modeling of the problem

The neural network architecture used to model the equation of conformal fractional isothermal gas spheres is shown in Fig. 1. Write Equation (2) as

$$D_x^{\alpha\alpha} u + \dfrac{1}{x^{2\alpha}} D_x^\alpha u = e^{-u}. \qquad (11)$$

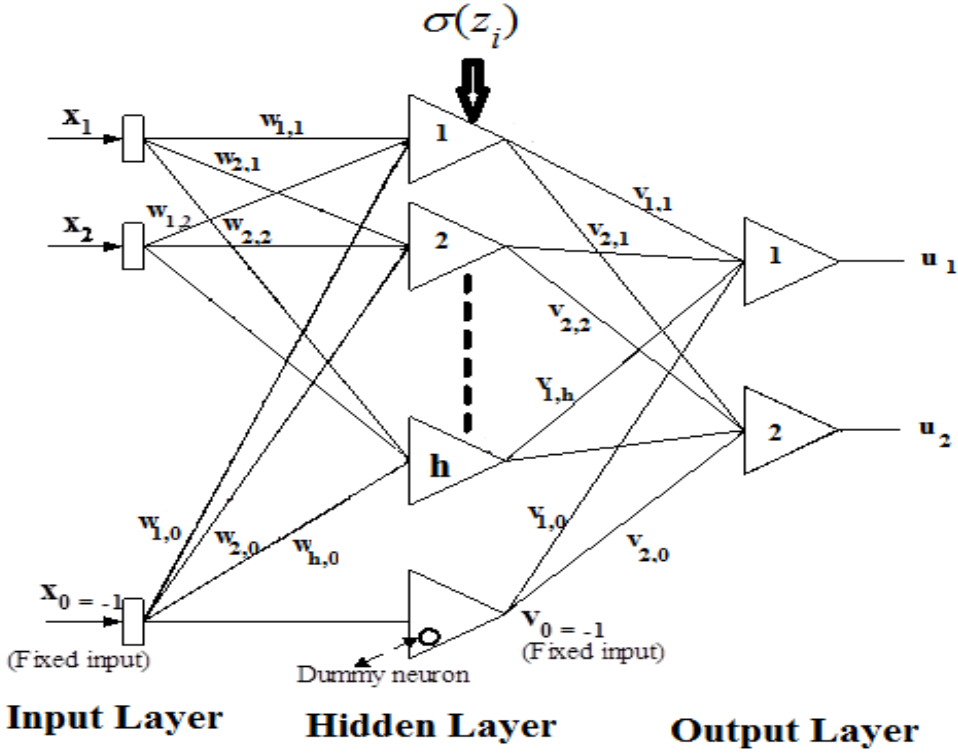

**Fig. 1. ANN Architecture developed to simulate the fractional isothermal Emden function, mass-radius relation, and density profiles.**

Along with the initial conditions $u(0) = 1$ and $D_x^\alpha u(0) = 0$, we receive a neural network solution, and we go through the following scheme:

First, we suppose that the solution of Equation (11) is $u_t(x, p)$ which can be approximated by

$$u_t(x,p) = A(x) + f(x, N(x,p)), \tag{12}$$

where $A(x)$ fulfills the initial conditions and $f(x, N(x,p))$ indicates the feed-forward neural network and $N(x, p)$ is the output of the neural network. The vector $x$ is the network input and $p$ is the analogous vector of adaptable weight parameters. Then $N(x, p)$ could be written as

$$N(x,p) = \sum_{i=1}^{H} v_i \sigma(z_i), \tag{13}$$

where $z_j = \sum_{i=1}^{n} w_{ij} x_j + \beta_i$ and $w_{ij}$ represents the weight from unit $j$ in the input layer to the unit $i$ in the hidden layer, $v_i$ symbolizes the weight from unit $i$ in the hidden layer to the output, $\beta_i$ is the bias value of the *i*th hidden unit and $\sigma(z_i)$ is the sigmoid activation function which has the form $\sigma(x) = \dfrac{1}{1+e^{-x}}$.

Taking the fractional derivative $N(x,p)$ for input vector $x_j$ gives

$$D_{x_j}^{\alpha} N(x,p) = D_{x_j}^{\alpha}\left(\sum_{i=1}^{H} v_i \sigma\left(z_i = \sum_{i=1}^{n} w_{ij} x_j + \beta_i\right)\right) = \sum_{i=1}^{h} v_i w_{ij} \sigma^{(\alpha)}, \qquad \sigma^{(\alpha)} = D_x^{\alpha}\sigma(x), \qquad (14)$$

the $n^{th}$ fractional derivative of $N(x,p)$ gives

$$D_{x_j}^{\alpha \cdots \alpha \; n\;times} N(x,p) = \sum_{i=1}^{n} v_i \, P_i \, \sigma_i^{(n\alpha)}, \qquad , P_i = \prod_{k=1}^{n} w_{ik}^{\alpha_k}, \quad \sigma_i = \sigma(z_i), \qquad (15)$$

Then, the approximate solution is then given by

$$u_t(x,p) = x N(x,p), \qquad (16)$$

This fulifies the initial conditions as:

$$u_t(0,p) = 0.N(0,p) = 0, \qquad (17)$$

and

$$D_x^{\alpha} u_t(x,p) = x^{1-\alpha} N(x,p) + x D_x^{\alpha} N(x,p), \qquad (18)$$

so

$$D_x^{\alpha} u_t(0,p) = (0)^{1-\alpha} N(x,p) + 0.D_x^{\alpha} N(x,p) = 0, \qquad (19)$$

### 3.2. Gradient computations and parameter updating

Assuming that Equation (16) represents the approximate solution, the problem will be turned into an unconstrained optimization problem and the amount of error will be given by

$$E(x) = \sum_i \left\{ D_x^{\alpha\alpha} u_t(x_i, p) + \dfrac{2}{x^{\alpha}} D_x^{\alpha} u_t(x_i, p) - f(x_i, u_t(x_i, p)) \right\}^2, \qquad (20)$$

Where:

$$f(x_i, u_t(x_i, p)) = e^{-u_t(x_i, p)}, \quad D_x^{\alpha} u_t(x,p) = x^{1-\alpha} N(x,p) + x D_x^{\alpha} N(x,p), \qquad (21)$$

and

$$D_x^{\alpha\alpha} u_t(x,p) = (1-\alpha) x^{1-2\alpha} N(x,p) + 2x^{1-\alpha} D_x^{\alpha} N(x,p) + x D_x^{\alpha\alpha} N(x,p), \qquad (22)$$

where $D_x^{\alpha} N(x,p)$ and $D_x^{\alpha\alpha} N(x,p)$ are given by Equations (14, 15).

We computed the fractional derivative of the neural network input as well as network parameters to update the network parameters and use the optimized parameter values to train the neural network. We set up the network with the optimized network parameters after the training of the network and calculate $u_t(x,p)$ from $u_t(x,p) = x N(x,p)$.

The conformable fractional derivative is considered on a par with a feed-forward neural network $N$ with one hidden layer for each of its inputs, with the same weight values w and thresholds $\beta_i$ with each weight $v_i$ being exchanged with $v_i P_i$ where $P_i = \prod_{k=1}^{n} w_{ik}^{\alpha_k}$. Furthermore, the transfer function of each hidden unit is exchanged with the fractional derivative of the sigmoid function in the $n^{th}$ order. Consequently, with regard to the parameters of the original network, the conformable fractional gradient $N$ of the original network is

$$
\begin{aligned}
D_{v_i}^{\alpha} N &= P_i \; \sigma_i^{(n\alpha)} \\
D_{\beta_i}^{\alpha} N &= v_i P_i \; \sigma_i^{((n+1)\alpha)} \\
D_{w_{ij}}^{\alpha} N &= x_i v_i P_i \; \sigma_i^{((n+1)\alpha)} + v_i \alpha_j w_{ij}^{1-\alpha_j} \left( \prod_{k=1, k \neq j} w_{ik}^{\alpha_k} \right) \sigma_i^{(n\alpha)}
\end{aligned}
\quad (23)
$$

The updating rule of the network parameters can be specified as,

$$v_i(x+1) = v_i(x) + a D_{v_i}^{\alpha} N, \tag{24}$$

$$\beta_i(x+1) = \beta_i(x) + b D_{\beta_i}^{\alpha} N, \tag{25}$$

$$w_{ij}(x+1) = w_{ij}(x) + c D_{w_{ij}}^{\alpha} N, \tag{26}$$

where $a, b, c$ are learning rates, $i = 1, 2, \ldots, n$, and $j = 1, 2, \ldots, h$.

### 3.3. Back-propagation Training Algorithm

The back-propagation (BP) training algorithm is a gradient algorithm aimed to minimize the average square error between the desired output and the actual output of a feed-forward network. Continuously differentiable non-linearity is required for this algorithm. The gradient algorithm mathematics must assure that a specific node has to be adapted in a direct rate to the error in the units it is connected to. This algorithm has been described in detail in our previous paper Nouh et al. (2020). Fig. 2 shows a flow chart of an off-line back-propagation training algorithm, Nouh et al. (2020), Yadav et al. (2015):

The back-propagation (BP) learning algorithm is a recursive algorithm starting at the output units and working back to the first hidden layer. A comparison of the desired output $t_j$ with the actual output $u_j$ at the output layer is executed using an error function which has the following form:

$$\delta_j = u_j(t_j - u_j)(1 - u_j). \tag{27}$$

The error function for the hidden layer takes the following form:

$$\delta_j = u_j(1 - u_j)\sum_k \delta_k w_k. \tag{28}$$

where $\delta_j$ is the error term of the output layer, and $w_k$ is the weight between the output and hidden layers. The update of the weight of each connection is implemented by replicating the error in a backward direction from the output layer to the input layer as follows:

$$w_{ji}(t+1) = w_{ji}(t) + \eta \delta_j u_j + \gamma(w_{ji}(t) - w_{ji}(t-1)) \tag{29}$$

Learning rate $\eta$ is chosen such that it is neither too large leading to overshooting nor very small leading to a slow convergence rate. The last part in Equation (29) is the momentum term which is affixed with a constant $\gamma$ (momentum) to accelerate the error convergence of the back-propagation learning algorithm and also to assist in pushing the changes of the energy function over local increases and boosting the weights in the direction of the overall downhill, Denz (1998). This part is used to add a portion of the most recent weight values to the current weight values. The values of η and γ terms are set at the beginning of the training phase and determine the network speed and stability, Basheer and Hajmeer (2000).

The process is repeated for each input pattern until the output error of the network is decreased to a pre-specified threshold value. The final weight values are frozen and utilized to get the precise fractional values of isothermal gas spheres differential equations during the test phase. The quality and success of training of ANN are assessed by calculating the error for the whole batch of training patterns using the normalized RMS error that is defined as:

$$E_{rms} = \frac{1}{PJ}\sqrt{\sum_{p=1}^{P}\sum_{j=1}^{J}(t_{pj} - u_{pj})^2} \tag{30}$$

where $J$ is the number of output units, $P$ is the number of training patterns, $t_{pj}$ is the desired output at unit $j$, and $u_{pj}$ is the actual output at the same unit $j$. A zero error denotes that all the output patterns computed by the isothermal gas spheres ANN perfectly match the values expected and that the isothermal gas spheres ANN is fully trained. Similarly, internal unit thresholds are adjusted by supposing they are connection weights on links from the input with an auxiliary constant-value.

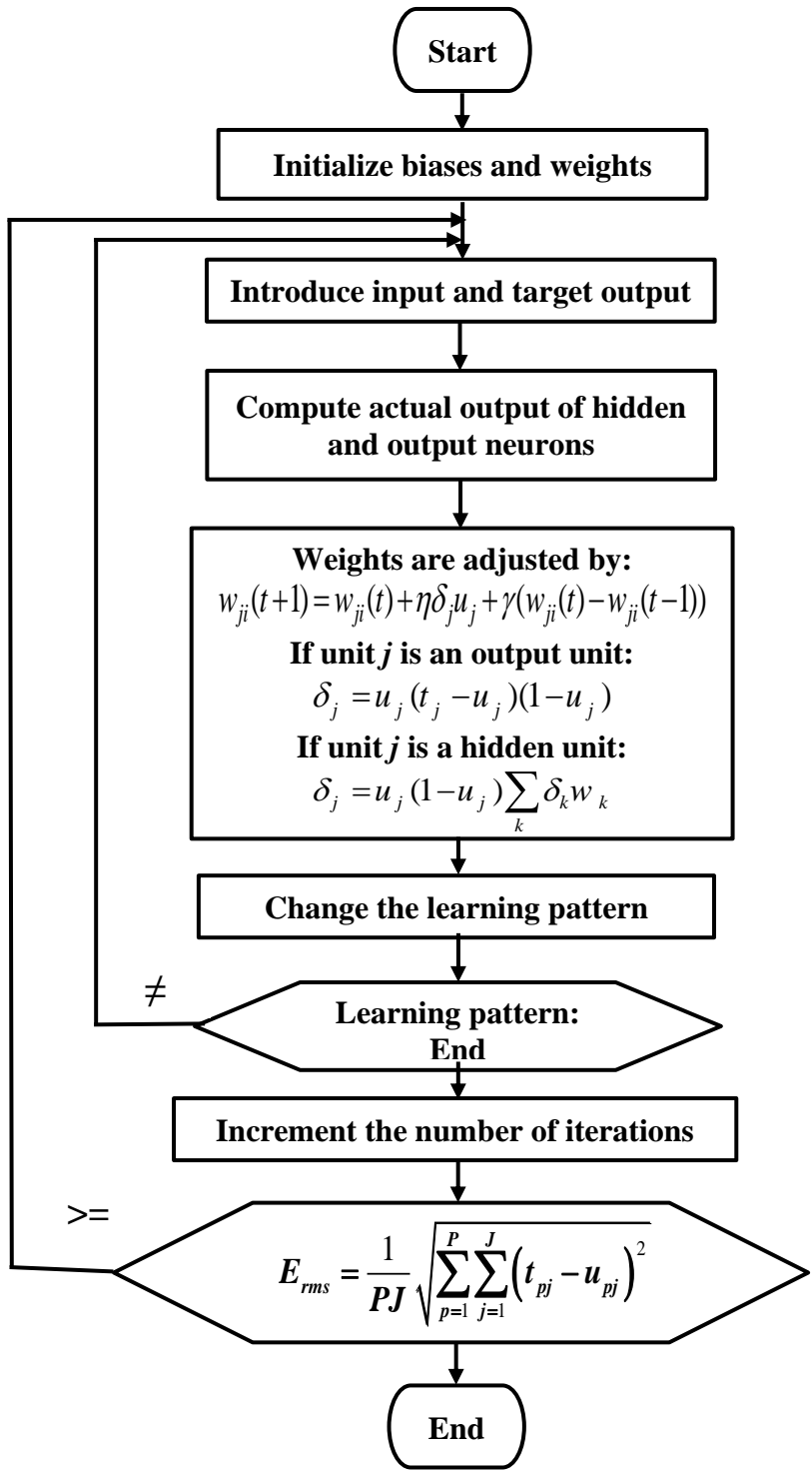

**Fig 2. Flowchart of an off-line back-propagation training algorithm**

# 4. Results and discussions
## 4.1 Data Preparation

We developed a MATHEMATICA routine to calculate the fractional Emden function and the physical characteristics of the conformable isothermal gas spheres like mass (Equation (3)), radius (Equation (4)), and density (Equation (5)). Equation (10) represents the series solution of fractional isothermal gas spheres which is similar to the power series solution developed by Yousif et al. (2020), where we used only 10 series terms. As pointed out by Yousif et al. (2020), this series expansions (like Equation (10)) diverge for $x \succ 3.2$. We used the accelerated scheme developed by Nouh (2004) to accelerate the series. Our calculations are done for $x \leq 80$ and the range of the fractional parameter $0.75 \leq \alpha \leq 1$ with step 0.1. For the integer case ($\alpha = 1$), the comparison between the Emden function computed according to the series solution and the numerical one is in good agreement, Yousif et al. (2021). Fractional models for the isothermal gas sphere could be computed using Equations (3) to (5) for the mass, radius, and density. So, we can investigate the mass-radius relations and density profiles at different fractional parameters. In Table 1, we listed the mass-radius relations and density profiles for some fractional isothermal gas spheres models. The designation in the table is: $R_*$ and $M_*$ represent the radius, and mass of the fractional star, $R_0$ and $M_0$ are the radius and mass of the sun. As appeared from the table, as the fractional parameter increases the volume and masse of the star decreases.

**Table 1: Mass-radius relations and density profile for the fractional isothermal gas sphere.**

| $\alpha$ | $R_* / R_0$ | $M_* / M_0$ |
|---|---|---|
| 1 | 1 | 1 |
| 0.99 | 0.956 | 0.915 |
| 0.98 | 0.915 | 0.838 |
| 0.97 | 0.875 | 0.768 |
| 0.96 | 0.838 | 0.703 |
| 0.95 | 0.802 | 0.644 |
| 0.94 | 0.767 | 0.591 |
| 0.93 | 0.735 | 0.514 |
| 0.92 | 0.703 | 0.495 |
| 0.91 | 0.673 | 0.464 |
| 0.9 | 0.644 | 0.415 |

**4.2 Network training**

To train the proposed neural network used to simulate the conformable fractional isothermal gas spheres equation we used data calculated in the previous subsection. The data used for training of the ANN are as shown in the second column of Tables (2-3).

**Table 2: Training, validation, and testing data for the fractional isothermal Emden function**

| Training phase | Validation phase | Testing phase |
|---|---|---|
| α | α | α |
| 0.8, 0.85, 0.9, 0.95, 0.97, 0.98, 1 | 0.96, 0.99 | 0.91, 0.92, 0.93, 0.94 |

**Table 3: Training, validation, and testing data for mass-radius relations and density profiles.**

| Training phase | Validation phase | Testing phase |
|---|---|---|
| α | α | α |
| 0.75, 0.90, 0.92, 0.93, 0.94, 0.96, 0.97, 0.99 | 0.95, 0.98 | 0.80, 0.85, 0.91 |

The architecture of the neural network (NN) we used in this paper for the isothermal gas sphere function is 2-120-1, where the input layer has two inputs which are the fractional parameter α and the dimensionless parameter $x$ ($x$ takes values from 0 to 80 in steps of 0.1), while the output layer has 1 node for the isothermal gas sphere function $u$ computed for the same values of the dimensionless parameter $x$ and input fractional parameter. For the mass-radius relation, we used the architecture 2-120-2, where the input layer has two individual inputs which are the fractional parameter α and the radius of the star $R$, while the output layer, has 2 nodes which are the mass and density at the same values of the input fractional parameters.

The choice of 120 neurons in the hidden layer of the NN was decided according to the findings we reached in our previous research (Nouh et al., 2020) after testing 80,120 and 200 neurons in one hidden layer of NN (shown in Fig. 1), which gave the least RMS error and the best model for the network compared to the other two configurations for both isothermal and mass-radius relation cases.

After multiple modifications and adjustments to the parameters of the NN, it is converged to an RMS error value of 0.00002 for the training of the isothermal case and value of 0.000025 for training of the mass-radius relation and density profile case. During the raining of the NN, we used a value for the learning rate ($\eta = 0.03$) and for the momentum ($\gamma = 0.5$). Those values for the learning rate and momentum were proved to quicken the convergence of the back-propagation training phase without exceeding the solution. For the demonstration of the convergence and stability of the values computed for weight parameters of network layers, the behaviors of the convergence of the input layer weights, bias and output layer weights ($w_i$, $\beta_i$ and $v_i$) for the isothermal gas sphere case are as displayed in Fig. 3. As well, the convergence behaviors and stability of the values computed for weight parameters of network layers (weights of the input layer, bias, and output layer) for the mass-radius relation case are shown in Fig. 4. As these figures indicate, the values of the weights are initialized to random values and after somewhat considerable iterations they converged to stable values.

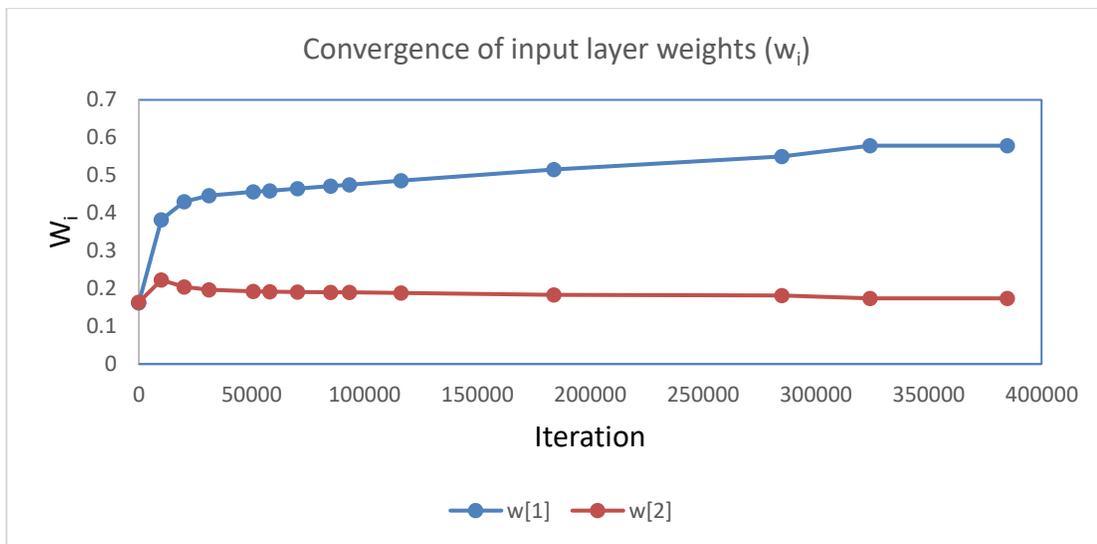

(a) The convergence of weights of the input layer ($w_i$)

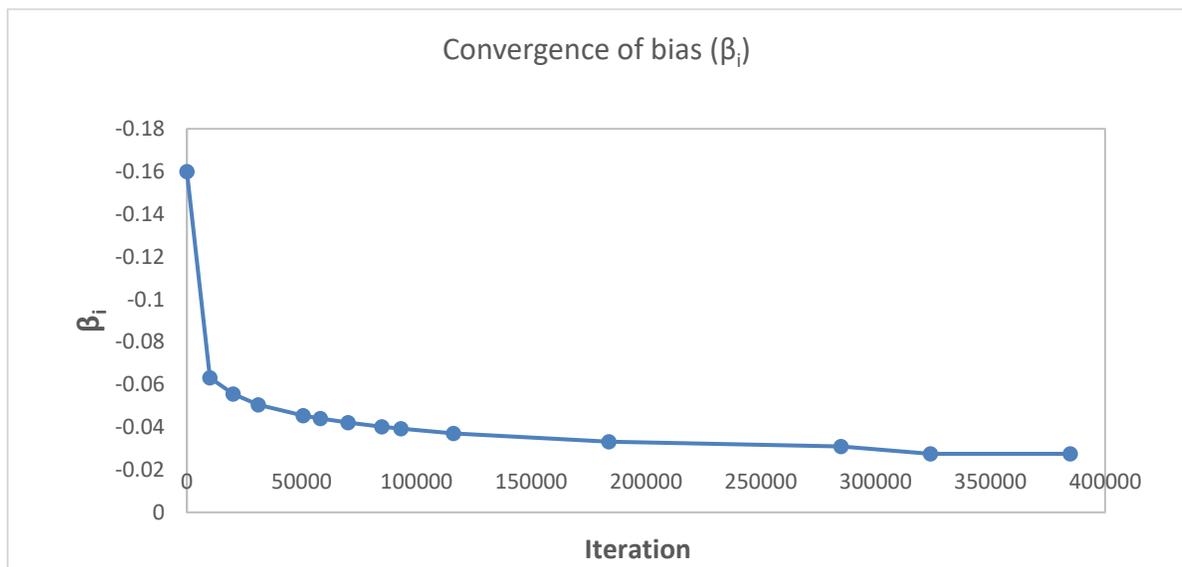

(b) The convergence of bias ($\beta_i$)

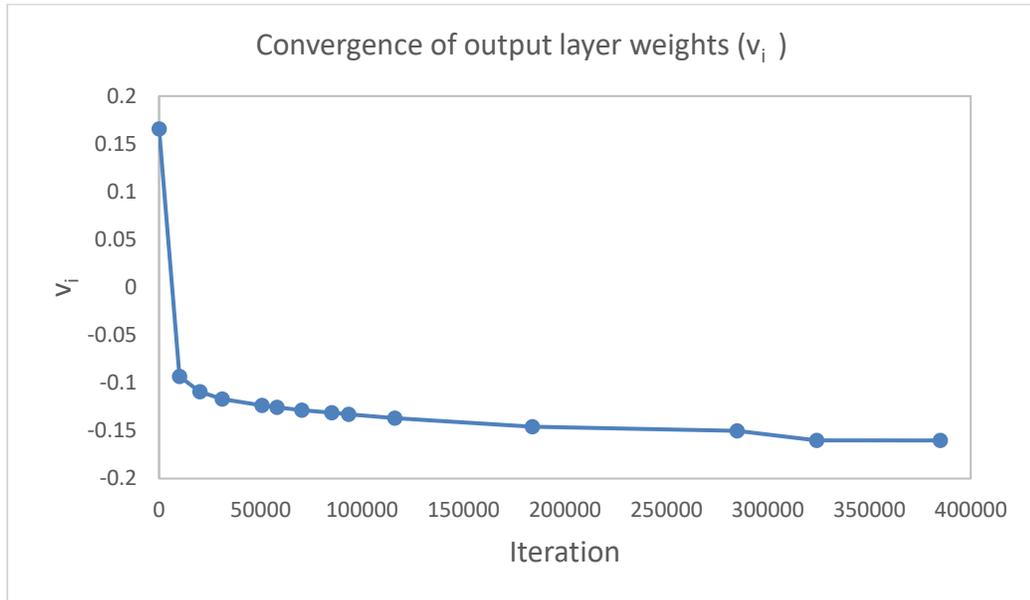

(c) The convergence of output layer weights ($v_i$)

**Figure (3) Convergence of input, bias, and output weights for the fractional isothermal Emden function.**

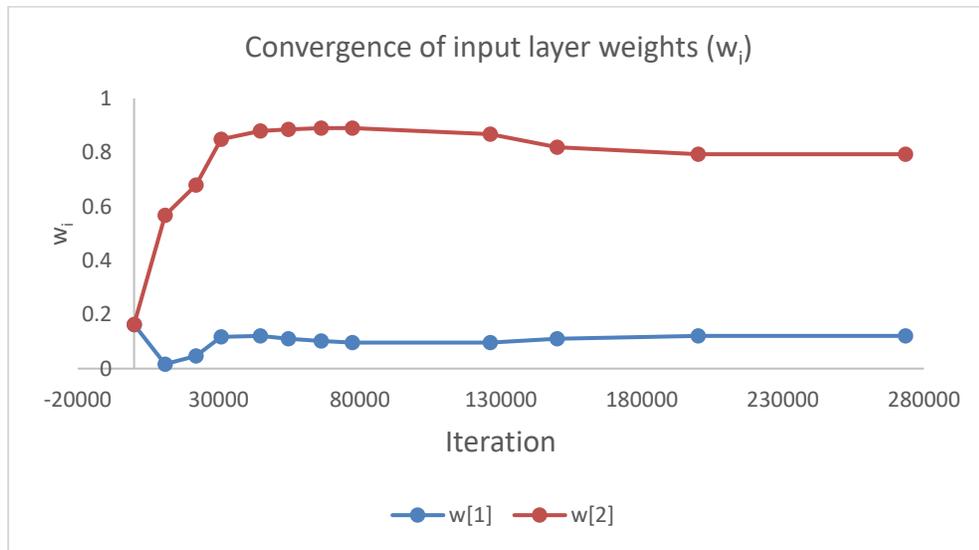

(a) The convergence of input layer weights ($w_i$)

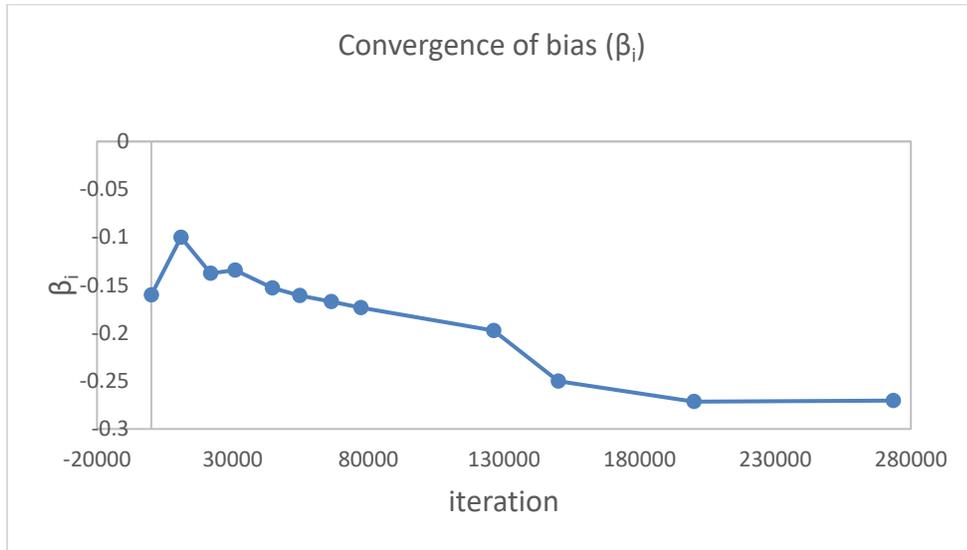

(b) The convergence of bias ($\beta_i$)

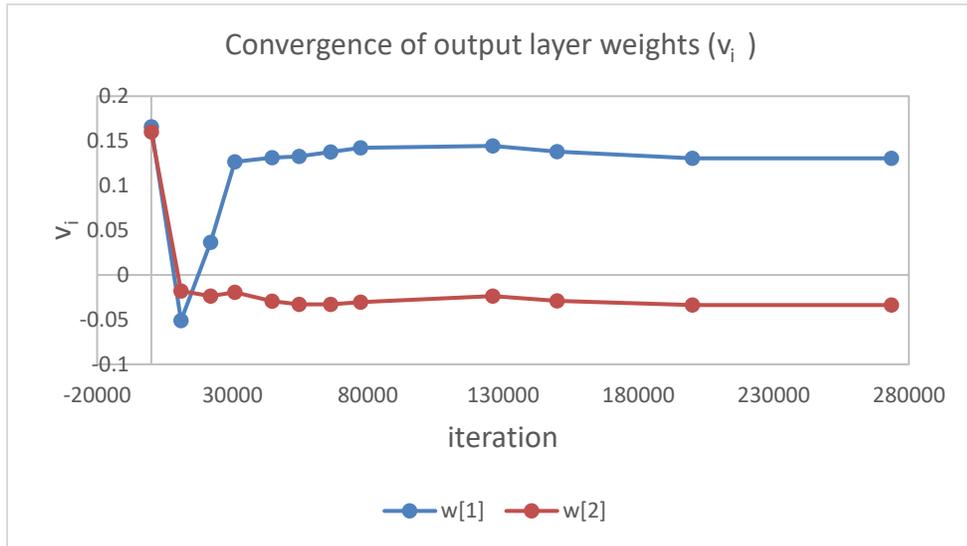

(c) Convergence of output layer weights ($v_i$)

**Figure 4: Convergence of the weights of input, bias, and output layers for the fractional mass-radius relation.**

### 4.3 Validation and Test of the Training phase

To ensure the training of the NN, we used two values for the fractional parameter α not being used in the training phase as a validation and verification of the goodness of that phase. These two values are shown in the middle column of Table 2 and Table 3 for the isothermal Emden function, and mass-radius relations and density profiles cases respectively. The obtained results for those two validation values are as shown in Fig. 5 for the isothermal Emden function, Fig. 7 for the fractional mass-radius-relations, and Fig.9 for the fractional density profiles. As are shown in these figures, there is a very good coincidence between the NN prediction and the analytical results for the Emden function, mass-radius relations, and density profiles, where the maximum absolute error is 1 %, 2.5 %, and 4 % respectively. We plotted the analytical solution and the NN prediction for the Emden function and the density profile with different colors, but due to the overlapping of the two curves, they appear as one. The big difference comes from the region near the center of the sphere, for $x \leq 10$. In the case of the mass-radius relation (Fig. 9), the noticeable difference between the analytical solution and the NN is larger than that of the Emden function and density profiles due to the nature of the equation related to the mass with radius (Equation (3)).

In Figures (6), (8), and (10), we plotted the predicted values of Emden functions, mass-radius relations, and the density profiles for some values of the fractional parameters listed in Table (1) and Table (2). In these figures, due to the small change of the Emden function and density with the fractional parameter and also the negligible difference between the analytical solution and the NN solution we truncated the x-axis at a smaller value for more clarity. Again, there is a somewhat noticeable difference between the analytical solution and the predicted NN values in the case of the mass-radius relation (Fig. 10) which is larger than the other two predicted NN values for Emden functions and the density profiles (Fig. 6 and Fig. 8). This large difference may be attributed to the instability during the performing and accelerating the series expansion of the fractional derivative of the Emden function (Equation (3)).

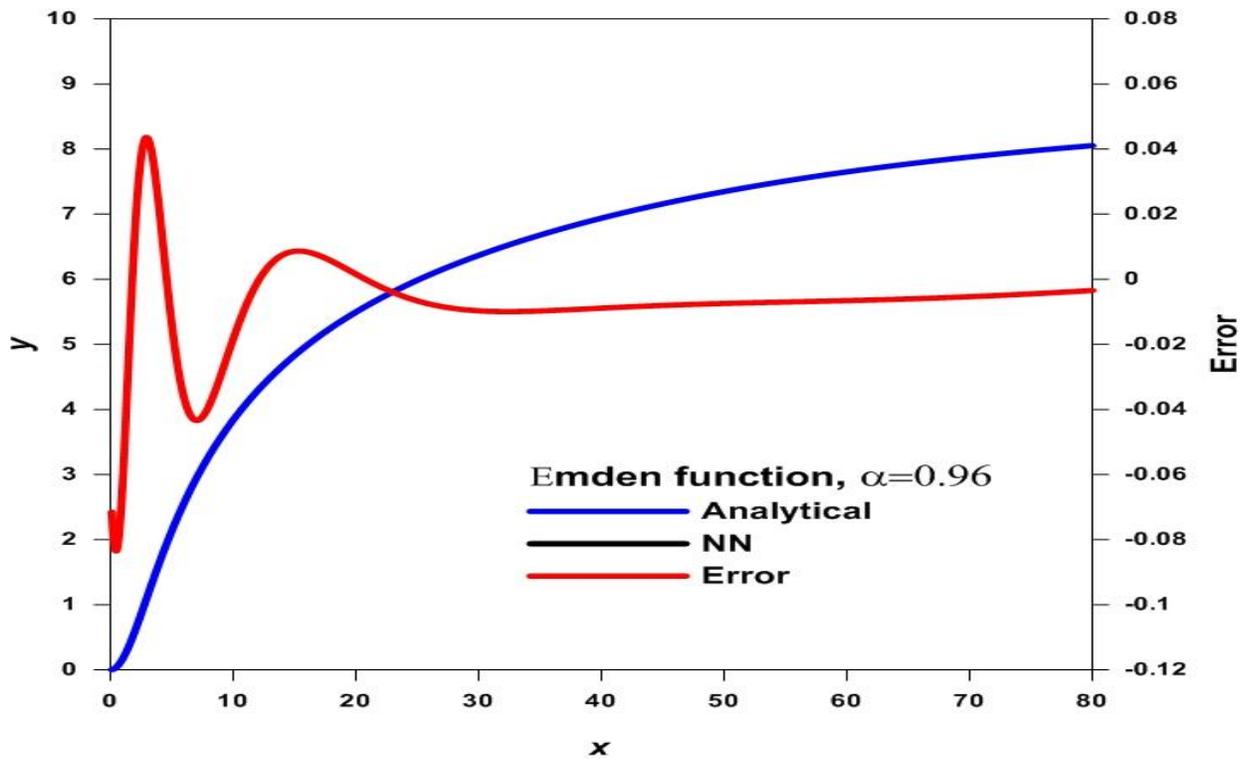

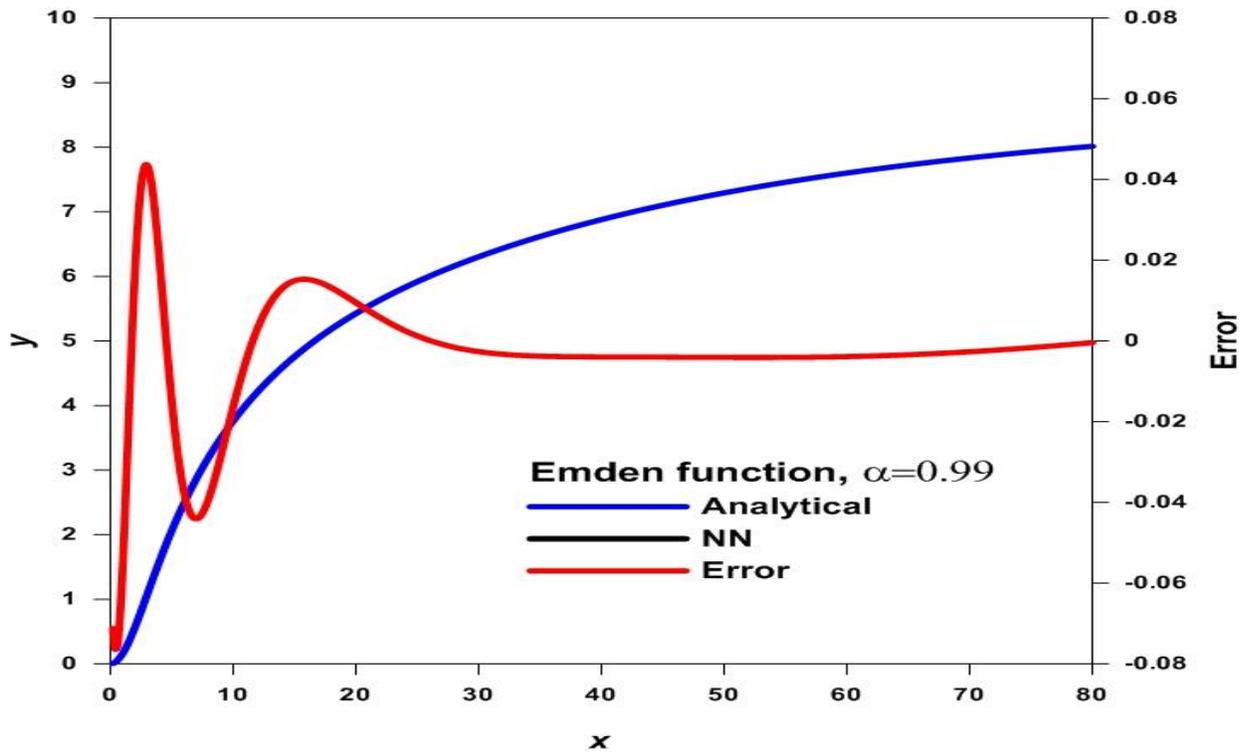

**Figure 5: The fractional Emden functions of the isothermal gas sphere obtained in the validation phase. The maximum relative error is 1%.**

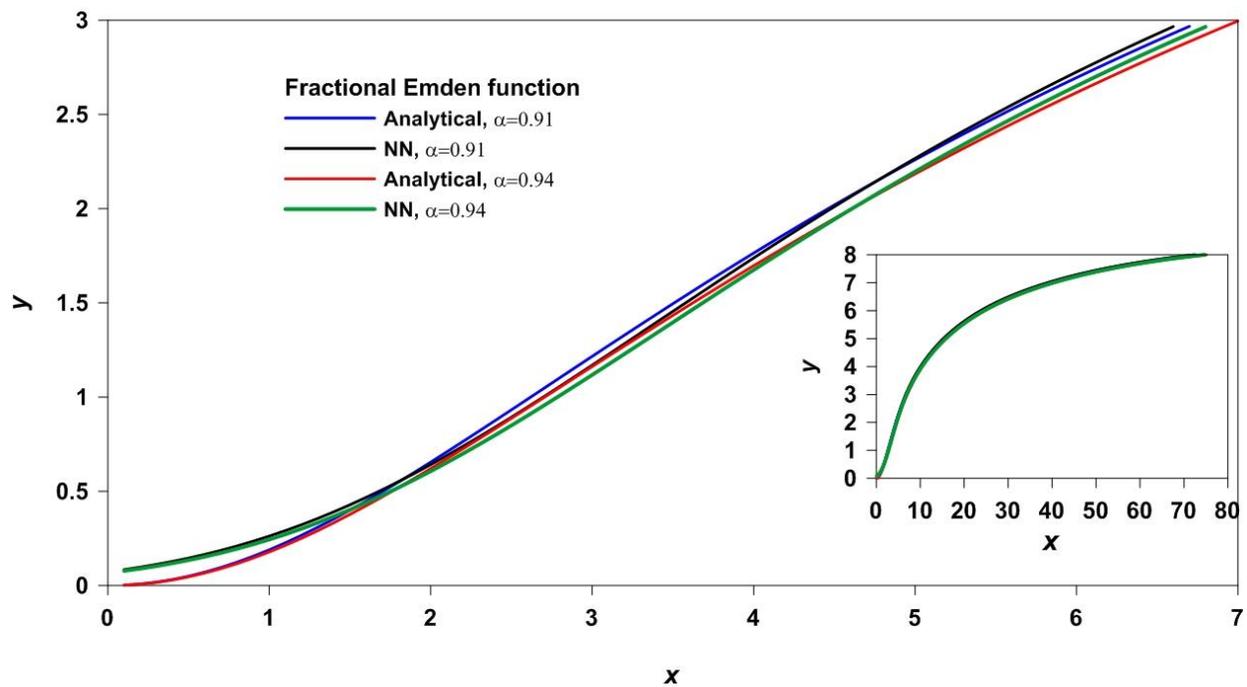

**Figure 6: The fractional Emden functions of the isothermal gas sphere obtained in the test.**

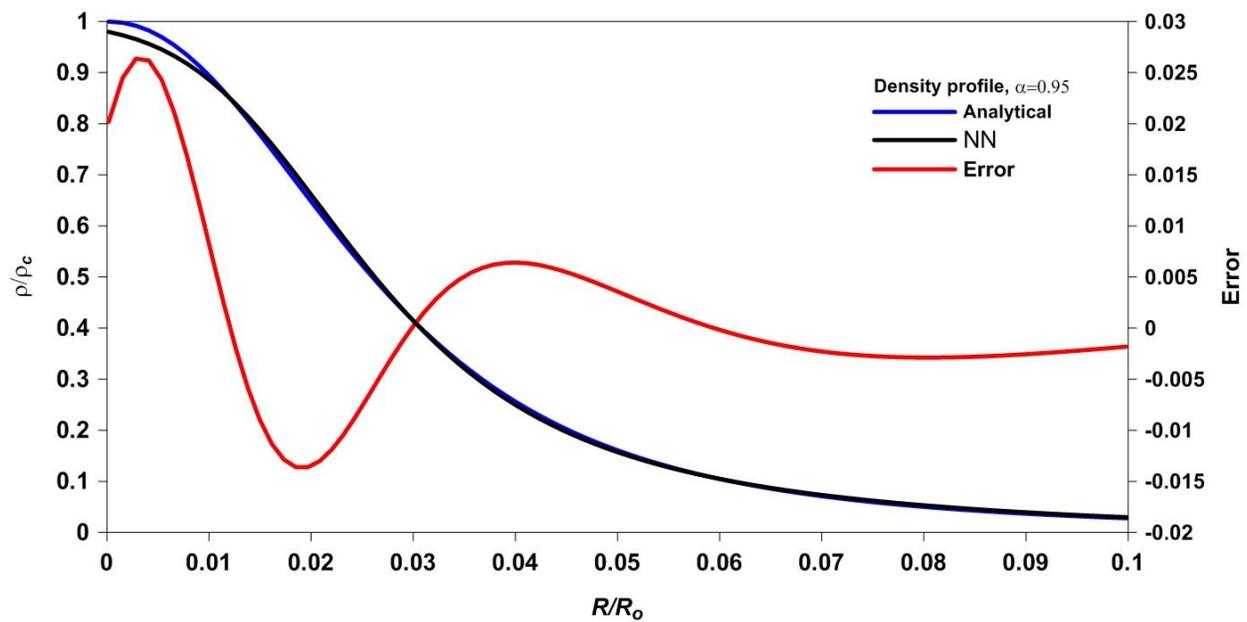

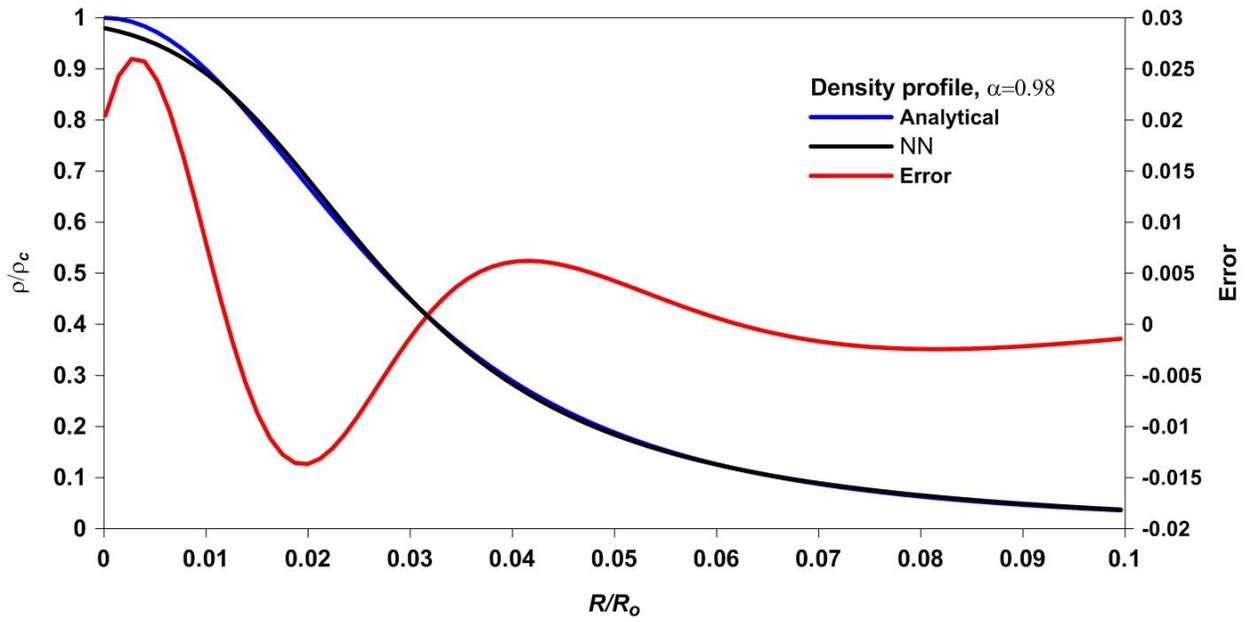

Figure 7: The fractional density profiles obtained in the validation phase. The maximum relative error is 2.5%.

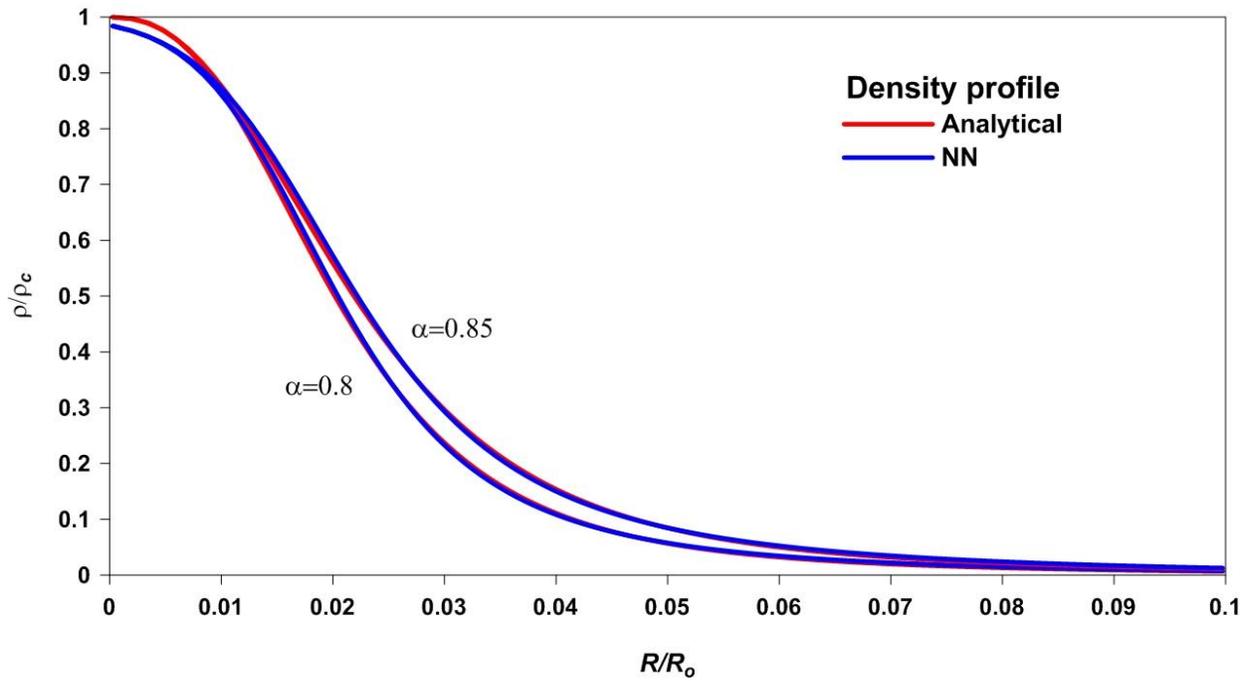

Figure 8: The fractional density profiles obtained in the test phase.

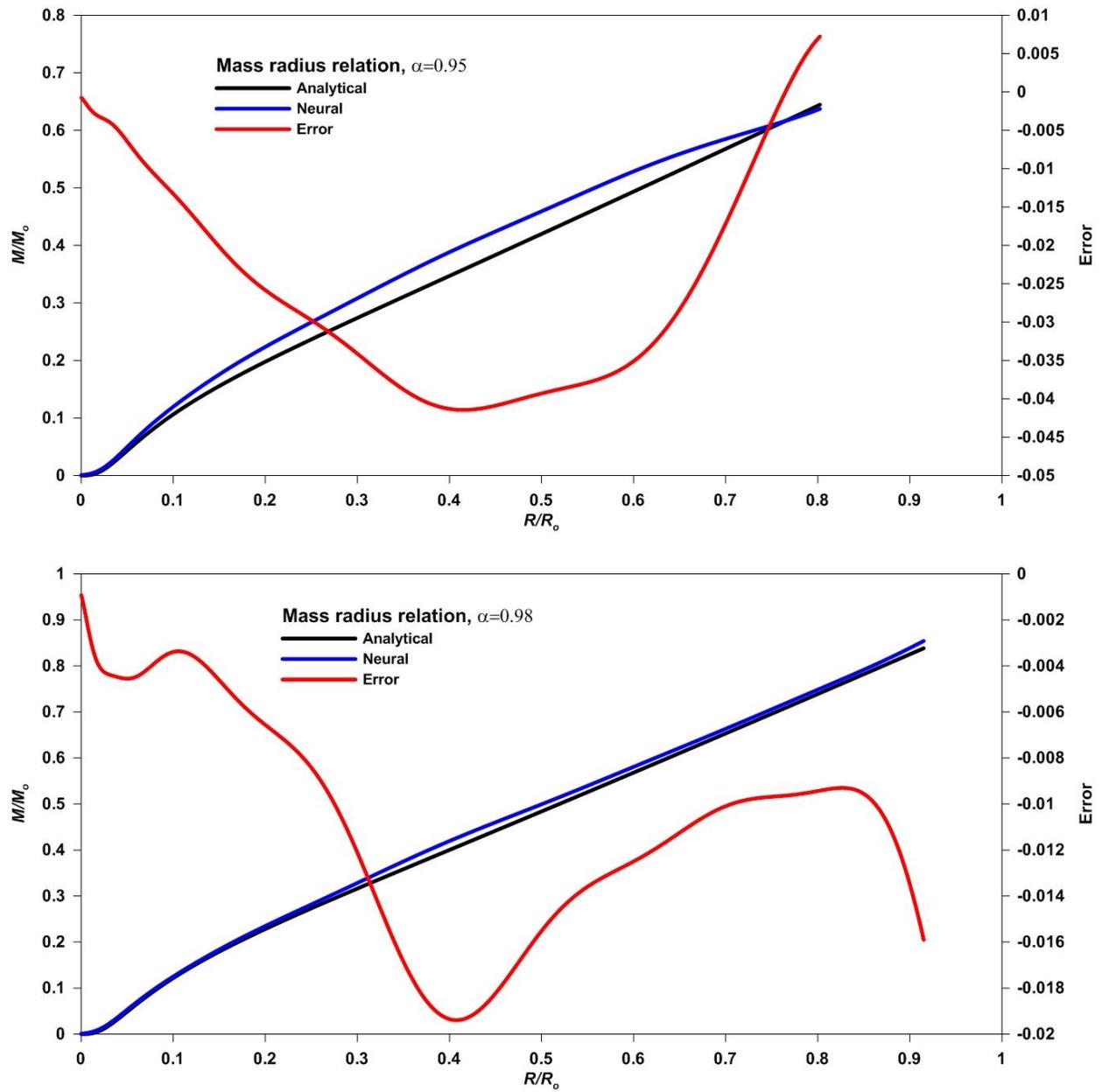

**Figure 9:** The fractional mass-radius relations obtained in the validation phase. The maximum relative error is 4%.

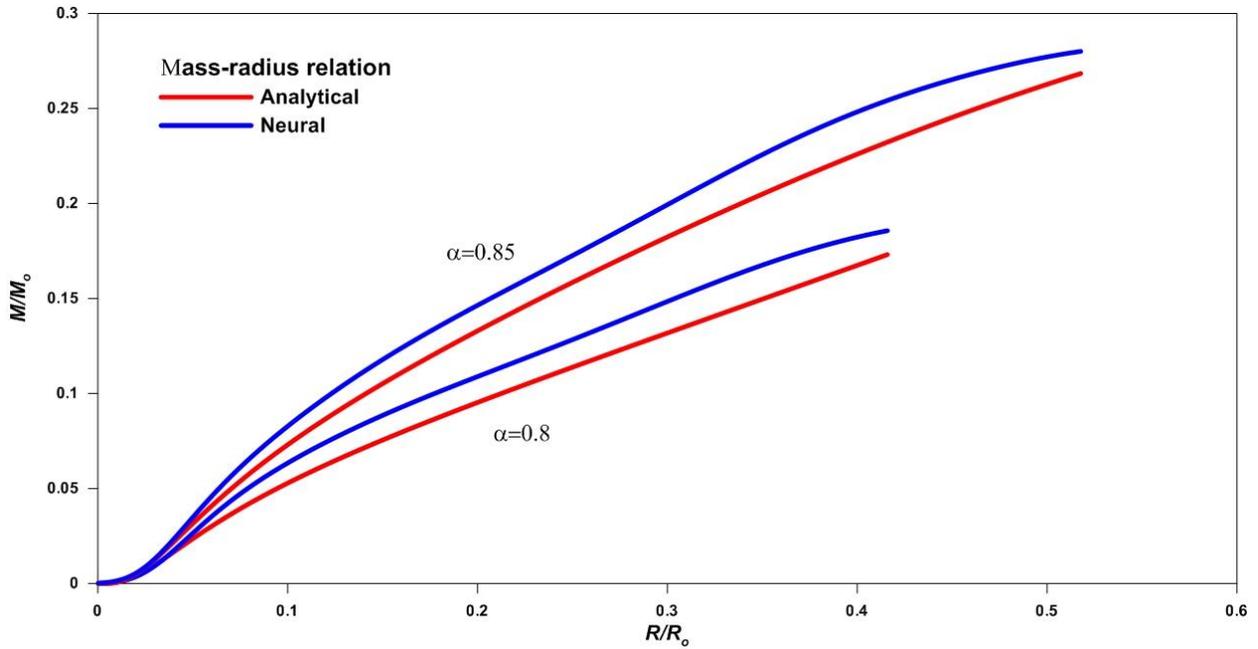

**Figure 10: The fractional mass-radius relations obtained in the test phase.**

## 5. Conclusion

The ANN modeling of the nonlinear differential equations proves high efficiency when compared with the numerical and analytical methods. In the present work, we aimed to introduce a computational approach to the fractional isothermal gas sphere via ANN. We solved the second type of Lane-Emden equation (the isothermal gas sphere) using the Taylor series, then we accelerated the resulted series to reach good accuracy. The analytical calculations are performed for the Emden functions, mass-radius relations, and density profiles.

We obtained good accuracy through the use of the ANN technique by using some calculated data to train the NN in the training phase, then validating the trained network by some other values, where we got maximum error values of 1 %, 2.5 %, and 4 % for isothermal fractional Emden function case, density profile case and mass-radius relation case respectively. To test the ANN technique in predicting unseen values, we used the trained network and ran the routine for the fractional test parameters listed in Tables 2 and 3. The comparison between the analytical and the ANN solution gives very good agreement as appeared in Figures (6, 8, and 10) with a maximum error of 4%. The results obtained reflect the applicability and efficiency of using ANN to model stellar physical characteristics (i.e. radius, mass, and density) using the fractional isothermal gas sphere. We think the present results besides the results obtained in Nouh et al. (2020) are an important step toward the composite modeling (e.g. isothermal core and polytropic envelope) of various stellar configurations using ANN.